\documentclass[aps,pra,twocolumn,superscriptaddress,showpacs]{revtex4}

\usepackage[latin1]{inputenc}
\usepackage[T1]{fontenc}
\usepackage{amsmath,amssymb}
\usepackage{bbm}
\usepackage{graphicx}
\usepackage{subfigure}
\usepackage{sistyle}

\ifx\pdfoutput\undefined
\else
\fi

\DeclareMathOperator{\diag}{diag}
\DeclareMathOperator{\var}{Var}
\DeclareMathOperator{\cov}{Cov}
\newcommand\relphantom[1]{\mathrel{\phantom{#1}}}

\begin{document}

\title{Gaussian State Description of Squeezed Light}

\author{Vivi Petersen}
\affiliation{QUANTOP - Danish National Research Foundation Center for
  Quantum Optics}
\affiliation{Department of Physics and Astronomy, University of
  Aarhus, DK-8000 Århus C, Denmark}
\author{Lars Bojer Madsen}
\affiliation{Department of Physics and Astronomy, University of
  Aarhus, DK-8000 Århus C, Denmark}
\author{Klaus Mølmer}
\affiliation{QUANTOP - Danish National Research Foundation Center for
  Quantum Optics}
\affiliation{Department of Physics and Astronomy, University of
  Aarhus, DK-8000 Århus C, Denmark}

\date{\today}

\begin{abstract}
  We present a Gaussian state description of squeezed light generated
  in an optical parametric oscillator. Using the Gaussian state
  description we describe the dynamics of the system conditioned on
  homodyne detection on the output field. Our theory shows that the
  output field is squeezed only if observed for long enough times or
  by a detector with finite bandwidth. As an application of the
  present approach we consider the use of finite bandwidth squeezed
  light together with a sample of spin-polarized atoms to estimate a
  magnetic field.
\end{abstract}


\maketitle

\section{Introduction}
\label{sec:introduction}

Quantum mechanical squeezing of optical fields represents a means to
improve precision measurements below the standard quantum noise limit
for optical detection. These measurements exercise a significant
back-action on the probed system, and to assess the achievement of a
detection scheme we need a formalism that can deal with light-matter
interaction and measurement induced state reduction continuously in
time. Now, a continuous wave beam of light is described by infinitely
many modes, for example in time or frequency domain, and the quantum
state of the light field and of the system interacting with the beam
is in general too complicated to be fully accounted for in a
Schr\"odinger picture representation. In many quantum optical problems
with constant or periodic driving Hamiltonians, it has been possible,
however, to provide solutions in the Heisenberg picture for the
relationship between Fourier transformed frequency components of the
field and system observables. Unfortunately this well-established
technique does not apply in conjunction with measurements on the joint
system acting locally in time and hence affecting all frequency
components of the observables at each measurement event. The present
paper paper does not provide a solution to this general problem.
Instead, we shall demonstrate that for a specific dynamics restricted
to a specific class of states, the so-called Gaussian states, a
significant reduction in the number of parameters needed to fully
characterize the system enables a complete description. The squeezed
light produced in an optical parametric oscillator (OPO) is in such a
Gaussian state, implying that the field is fully characterized by the
first-order and second-order correlation functions of
the field variables. The interaction with an atomic system may destroy
the Gaussian character, but we shall restrict our attention to optical
interaction with a large collection of atoms through an effective
collective atomic observable, which may in turn be well described by a
Gaussian quantum state. The Gaussian state
formalism~\cite{fiurasek02:_gauss_trans_distil_entan_gauss_states,giedke02:_charac_gauss_operat_distil_gauss_states,eisert03:_introd_basic_entan_theor_contin_variab_system2}
was recently
employed~\cite{molmer04:_estim_param_gauss_probes,madsen04:_spin_squeez_precis_probin_light2,petersen05:_magnet_entan_atomic_sampl}
for the off-resonant Faraday rotation-like interaction between a
continuous beam of light and an atomic ensemble. To describe the
interaction with a continuous wave of light, we
proposed~\cite{molmer04:_estim_param_gauss_probes,madsen04:_spin_squeez_precis_probin_light2,petersen05:_magnet_entan_atomic_sampl}
to treat the beam as a sequence of short segments of light incident on
the atoms. In the interaction, each light segment acquires some
entanglement with the atomic sample and causes a modification of the
atomic state when the light segment is probed after the interaction.
The description of the incident optical beam is simple if the state of
the field factorizes in components corresponding to each short segment
of the beam. This is indeed the case for a coherent state of light,
representing a normal laser beam. Realistic sources of squeezed light,
on the other hand, have a finite bandwidth of squeezing which implies
that correlations exist between the field observables at different
times. Here, we extend the Gaussian formalism to the case of the
continuous output from an optical parametric oscillator (OPO). We
reproduce the properties of the squeezed optical beam which are
already known from standard quantum optics treatments, and we apply
our formalism to the example of atomic magnetometry.

The paper is organized as follows. In Sec.~\ref{sec:gener-sque-light},
we recall some results of the standard treatment of squeezing in an
OPO. In Sec.~\ref{sec:gaussian-states}, we present the Gaussian state
description of this system. In Sec.~\ref{sec:magn-with-sque}, we turn
to magnetometry. The Larmor precession of an atomic sample caused by
an unknown magnetic field is probed by optical Faraday rotation, and
the value of the magnetic field is gradually determined as measurement
data are accumulated. The use of a squeezed light source improves the
magnetometer for probing times larger than the inverse bandwidth of
squeezing. In Sec.~\ref{sec:outlook}, we conclude.

\section{Generation of Squeezed Light in an optical parametric oscillator}
\label{sec:gener-sque-light}

In this section, we present a simple model of squeezed light
generation in a cavity with a non-linear medium which is pumped by a
classical pump beam at frequency $2\omega_c$, twice the cavity
resonance frequency, and giving rise to creation and annihilation of
pairs of photons by the Hamiltonian~\cite{scully97:_quant_optic} (we
use $\hbar=1$ throughout)
\begin{gather}
  \label{eq:14}
  \mathcal{H}_\text{int} = ig(a^{\dag2}-a^2) = g(x_cp_c+p_cx_c)
\end{gather}
where $a^\dag$ and $a$ are the creation and annihilation operators for
the light inside the cavity, and where the canonical conjugate
variables are
\begin{gather}
  \label{eq:45}
  \begin{split}
    x_c &= \tfrac{1}{\sqrt{2}} (a+a^\dag)\\
    p_c &= \tfrac{1}{i\sqrt{2}} (a-a^\dag).
  \end{split}
\end{gather}
We express the Hamiltonian in a frame rotating with the cavity
resonance frequency $\omega_c$, and consider the dynamics in this
rotating frame.

In the absence of losses, the Heisenberg equations of motion
\begin{gather}
  \label{eq:44}
   \begin{split}
    \dot{x}_c(t) &= 2gx_c(t)\\
    \dot{p}_c(t) &= -2gp_c(t)
  \end{split}
\end{gather}
can be solved straightforwardly, leading to an exponential squeezing
of the $p_c$-variable and an accompanying anti-squeezing of the
$x_c$-variable, which maintains a constant value of the uncertainty
product.

This model produces a squeezed state of a single light mode inside the
cavity, and such states are subject to detailed analysis in most text
books on quantum optics. Here, however, we aim at applications of
squeezed light and hence we are interested in the squeezing properties
of the light that leaks out of the cavity. This light propagates out
of the cavity into a continuous beam, which corresponds to a continuum
of modes in frequency space. We thus replace one of the perfectly
reflecting cavity mirrors with a mirror with a small transmittance,
which will lead to a loss of the cavity field with rate $\Gamma$. The
resulting intra-cavity field state can be found in many different
ways, but for our purpose it is sufficient to note that the cavity
mirror acts as a beam splitter for the intra-cavity field and for the
vacuum field ($x_\text{ph,in}, p_\text{ph,in}$) incident on the cavity,
see Fig.~\ref{fig:6}. At the partly transmitting mirror the incident
field is reflected into the output field. The output field is a linear
combination of the reflected incident field and the transmitted
intra-cavity field. Imagine an incident beam segment of duration
$\tau$, short enough that the intensity transmitted at the mirror and
the field amplitude built up by the Hamiltonian~\eqref{eq:14} can be
treated to lowest order in $\tau$. We can then iterate the Heisenberg
equations of motion for the intra-cavity field and the output field
from the cavity and we obtain
\begin{subequations}
  \label{eq:43}
  \begin{align}
    x_c(t+\tau)
    &= (\xi+2g\tau)x_c(t)
    + \sqrt{\Gamma\tau}x_\text{ph,in}(t)\label{eq:39}\\
    p_c(t+\tau)
    &= (\xi-2g\tau)p_c(t)
    + \sqrt{\Gamma\tau}p_\text{ph,in}(t)\label{eq:40}\\
    x_\text{ph,out}(t+\tau)
    &= -\sqrt{\Gamma\tau}x_c(t) + \xi x_\text{ph,in}(t)\label{eq:41}\\
    p_\text{ph,out}(t+\tau)
    &= -\sqrt{\Gamma\tau}p_c(t) + \xi p_\text{ph,in}(t).\label{eq:42}
  \end{align}
\end{subequations}
where $\xi^2 = 1-\Gamma\tau$ denotes the probability for the segment
to be reflected by the mirror. This quantity is very close to unity,
and consequently $\xi \approx 1-\Gamma\tau/2$. The
expressions~(\ref{eq:39}--\ref{eq:42}) are of course equivalent to the
ones obtained by the conventional input-output
formalism~\cite{collett84:_squeez_intrac_travel_wave_light,gardiner91:_quant_noise},
with the last terms in~(\ref{eq:39},\ref{eq:40}) having the
characteristic properties of Wiener noise increments in the limit of
small $\tau$. Since we assume that the input field is in the vacuum
state, Eqs.~(\ref{eq:39},\ref{eq:40}) can be solved directly for the
variances of the intra-cavity field quadratures, starting from the
vacuum at $t=0$, and taking the $\tau \to 0$ limit
\begin{align}
  \var(x_c) &= \frac{1}{2}
  \frac{\Gamma-4ge^{-(\Gamma-4g)t}}{\Gamma-4g}\label{eq:30}\\
  \var(p_c) &= \frac{1}{2}
  \frac{\Gamma+4ge^{-(\Gamma+4g)t}}{\Gamma+4g}.\label{eq:31}
\end{align}
If $4g<\Gamma$, we see that these equations approach steady state for
large times $t$. Since we are interested in operating the OPO in a
regime where steady state can be obtained, we assume from now on that
$4g<\Gamma$. The light inside the cavity is still squeezed as
expected, but it is entangled with the emitted light, and hence it is
not in a pure state and also not in a minimum uncertainty state.

In the conventional input-output description, by a Fourier
transformation to frequency space, the equations~\eqref{eq:43} become
algebraic equations, and the output field operators in frequency space
are expressed as linear combinations of the input operators at the
same frequencies but with frequency dependent
coefficients~\cite{walls94:_quant_optic}. All moments of the field
annihilation and creation operators have trivial expectation values in
the vacuum state. If $\omega$ denotes the difference between the
optical frequency and the cavity resonance frequency $\omega_c$, we
have for example the following expression for the normal ordered
expectation value of the output field when the system has reached
steady state (remembering $4g < \Gamma$)
\begin{gather}
  \label{eq:11}
  \langle: x(\omega), x(\omega') :\rangle
  = \frac{2\Gamma g}{(\frac{\Gamma}{2}-2g)^2+\omega^2}
  \delta(\omega+\omega').
\end{gather}
The Lorentzian frequency dependence implies a temporal correlation
between the light emitted at different times, which is due to the
common origin in the intra-cavity field. The field at a single
instance of time is obtained by a Fourier transformation of the
expressions in frequency space. This will involve all frequencies,
also the ones far from the cavity resonance and hence outside the
bandwidth of squeezing. Consequently, one will not observe squeezing
properties if one observes a light field in a time interval shorter
than $\sim 1/\Gamma$. Integrating the signal over a finite time
interval $T$, corresponding to detection of the variable
\begin{gather}
  \label{eq:24}
  x_T=\frac{1}{\sqrt{T}} \int_t^{t+T} x(t') dt'
\end{gather}
yields a quantity with normal ordered expectation value
\begin{gather}
  \label{eq:12}
  \begin{split}
    \langle: x_T^2 :\rangle
    &= \frac{1}{2T} \int_0^T \int_0^T :x(t)x(t'): \,dt \,dt'\\
    &= \frac{1}{4\pi T} \int_0^T \int_0^T \int_{-\infty}^\infty
    \int_{-\infty}^\infty :x(\omega)x(\omega'):\\
    & \hphantom{= \frac{1}{4\pi T} \int_0^T \int_0^T \int_{-\infty}^\infty
    \int_{-\infty}^\infty}
    e^{-i\omega t} e^{-i\omega't} \,d\omega \,d\omega' \,dt \,dt'\\
    &= \frac{8g\Gamma}{T(\Gamma-4g)^3}
    \left[ (\Gamma-4g)T - 2 + 2e^{(-\Gamma/2+2g)T} \right].
  \end{split}
\end{gather}
If we use that $\langle x_T^2 \rangle = \langle: x_T^2 :\rangle + 1/2$
we see that for short times ($(\Gamma-4g)T \ll 1$), the output field
has the standard noise of vacuum, whereas integration over a longer
time interval yields
\begin{gather}
  \label{eq:1}
  \var(x_T) \to \frac{1}{2} \frac{(\Gamma+4g)^2}{(\Gamma-4g)^2}.
\end{gather}
The corresponding variance for the $p_T$ component is obtained by
replacing $g$ by $-g$ in the above expressions, i.e., in the long-time
limit the emitted field is described by a minimum uncertainty state.

The prediction of the noise properties of $x_T$ and $p_T$ should of
course be in agreement with the ones observed if one carries out a
homodyne measurement to detect these quantities, but it is
important to remember that during such detection, the dynamics of the
system will be different, and it is not clear how to modify the
relations in frequency space between the intra-cavity and output
fields as the detection takes place in real time.

In the next section, we introduce the Gaussian state formalism which
allows an effective real-time treatment of the production \textit{and}
probing of squeezed light.

\section{Gaussian States}
\label{sec:gaussian-states}

\subsection{General Formalism}
\label{sec:general-formalism}

\begin{figure}[htbp]
  \centering
  \includegraphics{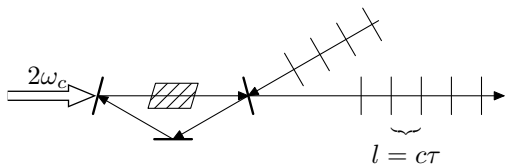}
  \caption{Generation of squeezed light by an optical parametric
    process pumped by a classical field at $2\omega_c$. The figure
    shows three field segments in the vacuum state which enter the
    cavity, where a non-linear medium generates squeezing. Four
    segments of light are shown propagating away from the cavity.}
  \label{fig:6}
\end{figure}
The linear transformation between the states of the cavity field and a
segment of light initially incident on the cavity, and eventually
propagating away from the cavity~\eqref{eq:43} is easy to deal with,
because a state which is initially Gaussian in the field variables
will remain Gaussian at later times. Taking an initially empty cavity
and the incident vacuum field, will hence lead to Gaussian states at
all later times. We want to consider the continuous emission of light
by the cavity, and we therefore imagine one segment of light after the
other leaving the cavity, see Fig.~\ref{fig:6}, and all field
quadratures being given by a multi-mode Gaussian distribution. A
Gaussian state is fully characterized by the mean value vector of all
canonical variables, which we arrange in a column vector $\mathbf{y}$,
with $\mathbf{m} = \langle\mathbf{y}\rangle$ and the covariance matrix
$\boldsymbol{\gamma}$ where $\gamma_{ij} = 2 \text{Re} \langle
(y_i-\langle y_i\rangle) (y_j-\langle y_j\rangle) \rangle$. If the
output field is discretized in $N$ segments $\mathbf{m}$ has dimension
$(2N+2)$ with $2N$ effective $x$ and $p$ variables for the output
field and 2 variables, $x_c$, $p_c$, for the cavity mode. The
covariance matrix $\boldsymbol{\gamma}$ has dimension $(2N+2) \times
(2N+2)$. These finite objects are of course far easier to deal with
than the full $N+1$ tensor products of infinite dimensional Hilbert
spaces. In practice the formalism can be made even simpler if we
assume that the output beam is detected right after it is emitted from
the cavity, and hence the quantum state of each light beam segment is
destroyed and only the classical output value is retained, while the
next segment emerges from the cavity. Let us consider the interaction
between a single incident segment of light and the intra-cavity field,
and let us write the linear transformation of the four field variables
$\mathbf{y} = (x_c,p_c,x_\text{ph},p_\text{ph})^T$ as follows
\begin{gather}
  \label{eq:17}
  \mathbf{y} \mapsto \mathbf{S}\mathbf{y}
\end{gather}
where the elements of the $4 \times 4$ matrix $\mathbf{S}$ follow
directly from the transformation~\eqref{eq:43}. Under this
transformation, the mean value vector $\mathbf{m}$ and the covariance matrix
$\boldsymbol{\gamma}$ transform as
\begin{align}
  \mathbf{m}(t+\tau)
  &= \mathbf{S} \mathbf{m}(t)\label{eq:18}\\
  \boldsymbol{\gamma}(t+\tau)
  &= \mathbf{S} \boldsymbol{\gamma}(t) \mathbf{S}^T.\label{eq:19}
\end{align}
We write the $4 \times 4$ covariance matrix as
\begin{gather}
  \label{eq:20}
  \boldsymbol{\gamma} =
  \begin{pmatrix}
    \mathbf{A}_\gamma & \mathbf{C}_\gamma\\
    \mathbf{C}_\gamma^T & \mathbf{B}_\gamma,
  \end{pmatrix}
\end{gather}
where $\mathbf{A}_\gamma$ is the covariance matrix for the
intra-cavity field variables, $\mathbf{B}_\gamma$ is the covariance
matrix for the beam segment in the continuous beam, and
$\mathbf{C}_\gamma$ represents their mutual correlations. An advantage
of the Gaussian description is that the back-action on the residual
system due to measurement may be accounted for explicitly. If we
measure the variable $x_\text{ph}$, due to their mutual correlation,
we learn something about the intra-cavity $x_c$ variable, i.e., its
variance decreases, and simultaneously, to fulfill Heisenberg's
uncertainty relation, $\var(p_c)$ increases. Following
Refs.~\cite{fiurasek02:_gauss_trans_distil_entan_gauss_states,giedke02:_charac_gauss_operat_distil_gauss_states,eisert03:_introd_basic_entan_theor_contin_variab_system2},
we have the explicit update formula for the intra-cavity field
covariance matrix after homodyne detection on the beam segment
\begin{gather}
  \label{eq:21}
  \mathbf{A}_\gamma
  \mapsto \mathbf{A}_\gamma
  - \mathbf{C}_\gamma
  (\pi\mathbf{B}_\gamma\pi)^- \mathbf{C}_\gamma,
\end{gather}
where $\pi = \diag(1,0)$, and $()^-$ denotes the Moore-Penrose
pseudoinverse. This result does not depend on the actual outcome of
the measurement. The latter however, affects the mean values of the
intra-cavity field variables. The beam segment has disappeared from
the treatment, but to treat the interaction with the next segment we
build the covariance matrix~\eqref{eq:20} describing the intra-cavity
field and this new segment with
\begin{align}
  \mathbf{B}_\gamma &\mapsto \mathbbm{1}\label{eq:22}\\
  \mathbf{C}_\gamma &\mapsto 0,\label{eq:13}
\end{align}
corresponding to an incident vacuum state with no correlation with the
cavity field yet. We propagate the system according to
Eq.~\eqref{eq:19}, and we implement the effect of the subsequent
measurement by Eq.~\eqref{eq:21}. The continuous production and
probing of the beam is obtained by repetition of the above steps, and
we may, in the limit of small time increments, derive a differential
equation for the intra-cavity field $\mathbf{A}_\gamma$. This
differential equation is of the general non-linear matrix Riccati form
(see, e.g., Ref.~\cite{stockton04:_robus_quant_param_estim} and
references therein)
\begin{gather}
  \label{eq:23}
  \dot{\mathbf{A}}_\gamma(t) = \mathbf{G}
  - \mathbf{D}\mathbf{A}_\gamma(t) - \mathbf{A}_\gamma(t)\mathbf{E}
  - \mathbf{A}_\gamma(t)\mathbf{F}\mathbf{A}_\gamma(t),
\end{gather}
where the $2 \times 2$ matrices $\mathbf{G,D,E,F}$ are all derived
from the expressions~\eqref{eq:19} and~\eqref{eq:21}. As shown in
Ref.~\cite{stockton04:_robus_quant_param_estim} the solution for
$\mathbf{A}_\gamma$ can be expressed in terms of the solutions of two
coupled linear matrix equations: $\mathbf{A}_\gamma =
\mathbf{W}\mathbf{U}^{-1}$, where $\dot{\mathbf{W}} =
-\mathbf{D}\mathbf{W} + \mathbf{G}\mathbf{U}$ and $\dot{\mathbf{U}} =
\mathbf{F}\mathbf{W} + \mathbf{E}\mathbf{U}$.

\subsection{Squeezing properties of the intra-cavity field}
\label{sec:intra-cavity-field}

Applying the above general formalism to the squeezed light problem, we
derive the Riccati equation~\eqref{eq:23} for the intra-cavity field
covariance matrix, conditioned on the homodyne detection of the output
field, and find the following matrices
\begin{gather}
  \label{eq:46}
  \begin{split}
    \mathbf{G} &= \left(
    \begin{smallmatrix}
      0 & 0\\
      0 & \Gamma
    \end{smallmatrix}\right)\\
    \mathbf{F} &= \left(
    \begin{smallmatrix}
      \Gamma & 0\\
      0 & 0
    \end{smallmatrix}\right)\\
    \mathbf{D} &= \left(
    \begin{smallmatrix}
      -2g-\Gamma/2 & 0\\
      0 & 2g+\Gamma/2
    \end{smallmatrix}\right)\\
    \mathbf{E} &= \mathbf{D}.
  \end{split}
\end{gather}

\begin{figure}[htbp]
  \centering
  \includegraphics[width=7cm]{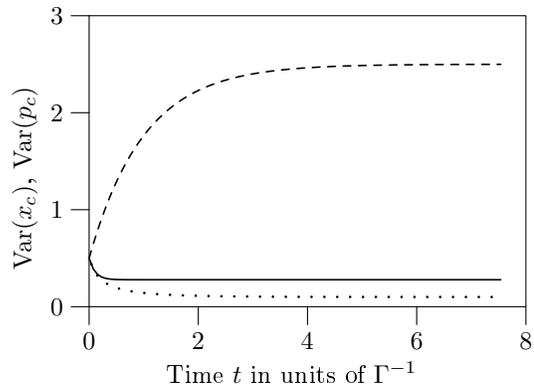}
  \caption{Variances of the cavity variables $x_c$ and $p_c$ as a
    function of time. We use $\Gamma = \SI{2\pi\times6e6}{}\ s^{-1}$
    and $g = 0.2 \Gamma$ which are realistic experimental parameters
    for OPO's~\cite{soerensen:_privat}. The variances of $x_c$ with
    and without homodyne detection of the $x_\text{ph}$ variable of
    the output field are identical and shown by the upper dashed
    curve. The full and the dotted curves show the variances of $p_c$
    without and with homodyne detection of the output field,
    respectively.}
  \label{fig:4}
\end{figure}

Without detection, the beam and the intra-cavity field are entangled,
and, as we noted above, the intra-cavity field state, regarded as a
trace over the unobserved emitted field degrees of freedom is a mixed
state and not a minimum uncertainty state. If we perform homodyne
detection on the emitted field we find the same variance of $x_c$ as
above, but the variance of $p_c$ changes to the value
\begin{gather}
  \label{eq:33}
  \var(p_c) = \frac{1}{2}
  \frac{\Gamma-4g}{\Gamma-4g e^{-(\Gamma-4g)t}}.
\end{gather}
In this case $\var(x_c)\cdot\var(p_c)=1/4$ and we have a minimum
uncertainty state of the intra-cavity field at all times. If
$4g<\Gamma$, we reach steady state for large times, and the variances
then read
\begin{align}
  \var(x_c) &= \frac{1}{2} \frac{\Gamma}{\Gamma-4g}\label{eq:34}\\
  \var(p_c) &= \frac{1}{2} \frac{\Gamma-4g}{\Gamma}\label{eq:35}.
\end{align}
In Fig.~\ref{fig:4}, we show how the variance of the Gaussian
variables inside the cavity depends on time both with and without
measurements on the output beam.

\subsection{Squeezing properties of the emitted beam}
\label{sec:sque-prop-emitt}

\subsubsection{Collective observable for many light segments}
\label{sec:coll-observ-many}

We now turn to the squeezing of the output beam. As discussed in
Sec.~\ref{sec:gener-sque-light}, there is no squeezing if we only
consider small time intervals. To study the correlations between
different individual segments we define the following collective
operators
\begin{gather}
  \label{eq:32}
  \begin{split}
    x_T &= \tfrac{1}{\sqrt{N}} \sum_{i=1}^N x_{\text{ph}_i},\\
    p_T &= \tfrac{1}{\sqrt{N}} \sum_{i=1}^N p_{\text{ph}_i},
  \end{split}
\end{gather}
where $T=\tau N$ is the accumulated time in $N$ segments each of
duration $\tau$ and where the field variables of the $i^\text{th}$
segment are retained in the formalism. In
appendix~\ref{sec:sque-outp-field}, we calculate the variances of
these quantities. The result is given in Eq.~\eqref{eq:10} and reads
\begin{gather}
  \label{eq:36}
  \begin{split}
    \var(x_T)
    &= \frac{1}{2T(\Gamma-4g)^3}
    \Bigl[ (\Gamma-4g)(\Gamma+4g)^2T\\
    &\relphantom{=} -32\Gamma g + 32\Gamma g e^{(-\Gamma/2+2g)T}\Bigr].
  \end{split}
\end{gather}
The result for $\var(p_T)$ is obtained by replacing $g$ with $-g$. If
we let $T \to 0$, we obtain $\var(p_T) = \frac{1}{2}$ showing that
there is no squeezing if we only consider short times. If, on the
other hand, we let $T \to \infty$, we obtain
\begin{gather}
  \label{eq:8}
  \var(x_T) \to \frac{1}{2} \frac{(\Gamma+4g)^2}{(\Gamma-4g)^2}.
\end{gather}
These results are in full agreement with the ones obtained by the
usual quantum optics treatment discussed in
Sec.~\ref{sec:gener-sque-light} (see Eq.~\eqref{eq:1}).

\subsubsection{Finite Bandwidth Detection}
\label{sec:detector-with-finite}

An alternative way to extract the squeezed component of the emitted
beam, is to use a frequency filter, that selects the frequency range
of interest. The modelling of such a detector involves a second
cavity, in which the light segments enter and the intra-cavity field
in the second cavity builds up. The squeezed beam contains photons in
the relevant frequency band, but not only intensity builds up in the
detecting cavity, we also expect the intra-cavity field to show
squeezing properties.
\begin{figure}[htbp]
  \centering
  \includegraphics[width=7cm]{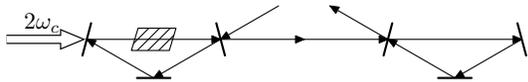}
  \caption{Proposed setup for the characterization of the spectrum of
    squeezed light from an OPO (to the left). The squeezing properties
    of the single-mode field accumulated in the frequency tunable
    cavity to the right are determined (see text).}
  \label{fig:13}
\end{figure}
The variables used in a Gaussian treatment of this problem,
corresponding to the two cavity fields and the propagating beam
segment, are $\mathbf{y} = (x_{c_1}, p_{c_1}, x_{c_2}, p_{c_2},
x_\text{ph}, p_\text{ph})$ and the Heisenberg equations of motion are
obtained by a simple extension of the expressions used already in the
case of a single cavity, where we replace $\Gamma$ with $\Gamma_1$.
The second cavity is used to model the finite bandwidth detection. It
has a decay constant $\Gamma_2$ and a tunable cavity resonance
frequency $\omega_c+\delta$. In our frame rotating at $\omega_c$ the
field variables in the second cavity obey the equations
\begin{align}
  x_{c_2}(t+\tau) &= (1-\Gamma_2\tau/2) x_{c_2}(t)
  + i\delta\tau p_{c_2}(t)
  + \sqrt{\Gamma_2\tau} x_\text{ph,out}(t)\label{eq:9}\\
  p_{c_2}(t+\tau) &= (1-\Gamma_2\tau/2) p_{c_2}(t)
  - i\delta\tau x_{c_2}(t)
  + \sqrt{\Gamma_2\tau} p_\text{ph,out}(t)\label{eq:15}
\end{align}
where $x_\text{ph,out}$, $p_\text{ph,out}$ are the quadrature
variables for the field leaving the first cavity, cf.,
Eqs.~(\ref{eq:41},\ref{eq:42}). In Eqs.~(\ref{eq:9},\ref{eq:15}) the
field incident on the second cavity is the output field from the first
cavity, cf. Fig.~\ref{fig:13}. Due to the physical separation $L$ of
the two cavities and the finite speed of light the field variables in
Eqs.~(\ref{eq:41},\ref{eq:42}) should in fact have been delayed by
$L/c$, but since we are addressing the steady state properties of the
system we can solve Eqs.~(\ref{eq:41},\ref{eq:42}) with the same time
arguments. The output field from the second cavity is described by
equations similar to Eqs.~(\ref{eq:41},\ref{eq:42}), but they will not
be needed in the following. The detuning $\delta$ of the second cavity
can be scanned, and the squeezing parameter of the intra-cavity
variables $x_{c_2}$, $p_{c_2}$ reflect the spectral properties of the
output beam from the first cavity.

\begin{figure}[htbp]
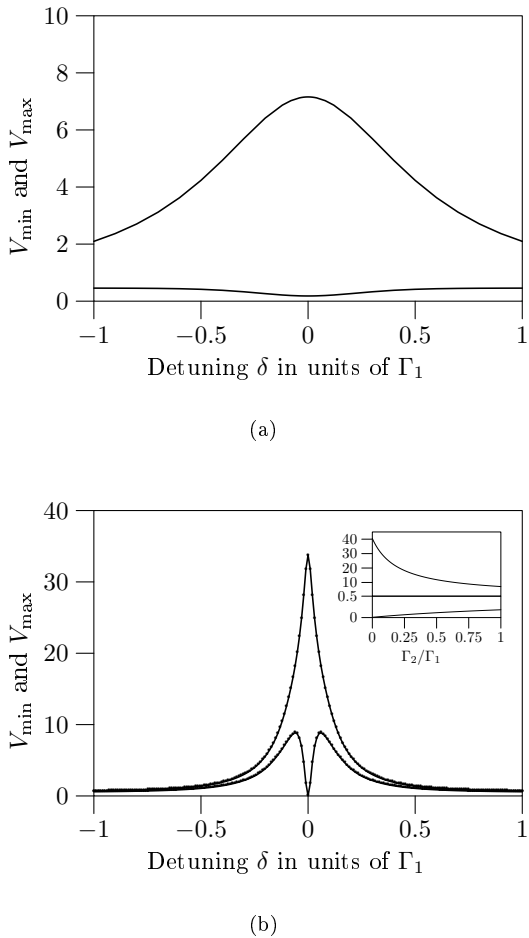

  \centering \subfigure[]{\label{fig:9}
    \includegraphics[width=7cm]{ViviFig4.eps}}\\
  \subfigure[]{\label{fig:11}
    \includegraphics[width=7cm]{ViviFig5.eps}}
  \caption{The variances $V_\text{min}$ and $V_\text{max}$ of the
    field inside the probing cavity as function of the detuning
    $\delta$ of this cavity with respect to $\omega_c$ in units of the
    decay width $\Gamma_1$ of the OPO cavity. We have used $\Gamma_1 =
    \SI{2\pi\times6e6}{}\ s^{-1}$ and $g = 0.2 \Gamma_1$ as in Fig.~2.
    In (a) $\Gamma_2 = \Gamma_1$ and in (b) $\Gamma_2 = \Gamma_1/25$.
    The lower and upper parts of the insert in (b) show $V_\text{min}$
    and $V_\text{max}$ respectively as functions of the bandwidth of
    detection $\Gamma_2$ for $\delta=0$.}
  \label{fig:11+9}
\end{figure}

Fig.~\ref{fig:11+9} shows the eigenvalues $V_\text{min}$ and
$V_\text{max}$ of the $2\times2$ covariance matrix for the probing
cavity as function of the detuning with respect to $\omega_c$. In
panel \ref{fig:9} the probing cavity has a damping rate $\Gamma_2$
comparable with the one of the OPO cavity, i.e., the intra-cavity
field builds up with a memory time shorter than the time needed to see
the full effect of squeezing. In panel~\ref{fig:11}, we use a detector
system with narrow bandwidth, the cavity builds up light over a longer
time interval, and the degree of squeezing is clearly larger than in
\ref{fig:9}. The Riccati equation can be solved analytically, and for
$\delta=0$ we obtain
\begin{gather}
  \label{eq:52}
  V_\text{max} = \frac{1}{2}
  \frac{(\Gamma_1+4g)^2+(\Gamma_1-4g)\Gamma_2}{
    (\Gamma_1-4g)(\Gamma_1+\Gamma_2-4g)}.
\end{gather}
Here $V_\text{min}$ is obtained from $V_\text{max}$ by replacing $g$
with $-g$. The insert shows $V_\text{min}$ and $V_\text{max}$ as
function of $\Gamma_2$. For large $\Gamma_2$, the second cavity is
equally fed by a wide range of frequency components, and the variance
is dominated by the vacuum uncertainty: $V_\text{min} = V_\text{max} =
1/2$. If $\Gamma_2=0$ we obtain
\begin{align}
  V_\text{max} &= \frac{1}{2}
  \frac{(\Gamma_1+4g)^2}{(\Gamma_1-4g)^2}\label{eq:53}\\
  V_\text{min} &= \frac{1}{2}
  \frac{(\Gamma_1-4g)^2}{(\Gamma_1+4g)^2}\label{eq:54}
\end{align}
which equal the long-time integrated amplitudes~\eqref{eq:8}.

We note that the calculations here were significantly easier than in
the case where we treated a large number of light segments
simultaneously. This is because the mode of the second cavity in
practice integrates the incident field over time and stores the
contribution of many short beam segments in a single set of variables.
We believe that this is a useful model of realistic finite bandwidth
detectors, and that the approach can be used quite generally to
investigate how finite optical bandwidth detection affects the
sensitivity of metrology and the entanglement and spin squeezing of
atomic samples.

\section{Magnetometry with Squeezed Light}
\label{sec:magn-with-sque}

The purpose of introducing the Gaussian state formalism is to provide
a theoretical approach, that allows a treatment of the interaction
between light and an atomic sample in the regime where the quantum
state of the atoms changes both because of the interaction itself and
because the continuous measurements of the light field after the
interaction teaches the observer about the state of the atoms. This
measurement induced back-action on the quantum state of atoms plays a
role in atomic
magnetometry~\cite{geramia03:_quant_kalman_filter_heisen_limit_atomic_magnet},
it was used to spin squeeze atomic gasses
\cite{kuzmich99:_quant_nondem_measur_collec_atomic_spin} and to
entangle pairs of gasses
\cite{julsgaard01:_exper_long_entan_two_macros_objec}, and it recently
played an important role in the realization of an atomic memory for
light~\cite{julsgaard04:_exper_demon_quant_memor_light}. Since probing
with squeezed light potentially is more precise, it was proposed in
Ref.~\cite{molmer04:_estim_param_gauss_probes} that magnetometry would
also benefit from the use of squeezed light, and a simple model with
ultra-broad band squeezing indeed suggests improvement by precisely
the squeezing factor on the B-field uncertainty. We will now use
magnetometry as an example to show how we can effectively treat the
probing of atomic systems with a real squeezed optical field with
finite bandwidth.

\begin{figure}[htbp]
  \centering
  \includegraphics{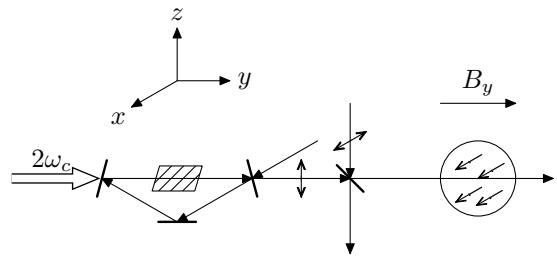}
  \caption{Setup for estimating a B-field. In the cavity, we generate
    squeezed light which is linearly polarized along the $z$~axis. We
    mix this field at an asymmetric beam splitter with a strong
    $x$~polarized beam. The light then passes through a gas of
    polarized atoms, causing a rotation of the field polarization
    towards the $z$~axis.}
  \label{fig:12}
\end{figure}

It is possible to estimate a magnetic field by a polarization rotation
measurement of an off-resonant light beam passing through a trapped
cloud of spin-$1/2$ atoms, see Fig.~\ref{fig:12}. All the atoms are
assumed to be polarized with their spin along the $x$~direction. We
assume that the B-field component of interest is directed along the
$y$~direction, and hence it causes a Larmor rotation of the atomic
spin toward the $z$~axis. This in turn leads to a mean magnetization
of the sample along the $z$ direction, which will cause a Faraday
rotation of the linear polarization of an optical field propagating
through the sample. As the polarization rotation is proportional to
the atomic spin component, and this is proportional to the B-field,
the B-field estimated by the measurement is trivially obtained. We
wish to address the error bar, i.e., the standard deviation on our
estimate of the field as a function of the measurement record. The
Gaussian state description which operates explicitly with the
variances and covariance elements of the physical quantities is ideal
for this analysis.

The gas of trapped spin-$1/2$ atoms is described by a collective spin
operator $\mathbf{J} = (1/2) \sum_i \boldsymbol{\sigma}_i$ where
$\boldsymbol{\sigma}_i$ are the Pauli spin matrices. The atoms are
initially pumped such that they are polarized along the $x$~axis and
$J_x$ can be treated as a classical variable $\langle J_x \rangle =
N_\text{at}/2$ where $N_\text{at}$ is the large number of atoms. The
two other projections of the spin, $J_y$ and $J_z$ obey the
commutation relation $[J_y, J_z] = i J_x$ which may be rewritten as
$[x_\text{at}, p_\text{at}] = i$ for the effective position and
momentum variables $x_\text{at} = J_y / \sqrt{ \langle J_x \rangle}$,
$p_\text{at} = J_z / \sqrt{\langle J_x \rangle}$. The uncertainty is
easily shown to be minimal in the initial state and, hence, the state
pertaining to $x_\text{at}$ and $p_\text{at}$ is Gaussian.

Note that the OPO cavity produces a squeezed vacuum state; if this
field is linearly polarized along the $y$-axis, it may be mixed on a
polarizing beam splitter with a classical $x$ polarized field to yield
the field appropriate for polarization rotation measurements.

The light beam propagates along the $y$~axis and is linearly polarized
along $x$ such that its Stokes operator $\langle S_x \rangle =
N_\text{ph}/2 = \Phi\tau/2$ is classical with $N_\text{ph}$ the number
of photons in a given segment of the beam and $\Phi$ is the photon
flux. The two remaining Stokes vector components corresponding to the
difference in photon numbers with linear polarization along directions
at 45 and 135 degrees with respect to the $x$~axis, and with left and
right circular polarizations, respectively, have vanishing mean
values, and they satisfy a commutator relation similar to the
collective atomic spin. Accordingly, for the effective variables
$x_\text{ph} = S_y / \sqrt{\langle S_x \rangle}$, $p_\text{ph} = S_z /
\sqrt{ \langle S_x \rangle}$, we have $[x_\text{ph}, p_\text{ph}] = i$
for $N_\text{ph} \gg 1$ and the initial coherent state of the field is
a minimum uncertainty Gaussian state in these variables.

The effective Hamiltonian for this part of the system is
\begin{gather}
  \label{eq:51}
  \mathcal{H}\tau = \kappa\sqrt{\tau} p_\text{at}p_\text{ph}
  + \mu\tau B_yx_\text{at}.
\end{gather}
The characteristic atom-light coupling is $\kappa =
\frac{d^2\omega}{\Delta Ac\epsilon_0} \sqrt{2\langle J_x \rangle
  \Phi}$ where $d$ is the atomic dipole moment, $\omega$ is the photon
energy, $\Delta$ is the detuning of the light from atomic resonance,
and $A$ is the area of the the light field. The coupling between the
B-field and the atoms is $\mu = \beta\sqrt{\langle J_x\rangle}$ where
$\beta$ is the magnetic moment.

The classical B-field that we wish to estimate is treated as a random
variable with a broad Gaussian probability distribution. Hence both
the B-field, the atomic cloud, and the incident light pulse of
duration $\tau$ are Gaussian variables. We arrange these in the vector
$\mathbf{y} = (B_y, x_\text{at}, p_\text{at}, x_\text{ph},
p_\text{ph})^T$. The Larmor precession in time $\tau$ and the
interaction between the atomic sample and the beam segment then leads
to the linear transformation~\eqref{eq:17} of the
variables~\cite{molmer04:_estim_param_gauss_probes,petersen05:_magnet_entan_atomic_sampl}
given by
\begin{gather}
  \label{eq:25}
  \mathbf{S} =
  \begin{pmatrix}
    1 & 0 & 0 & 0 & 0\\
    0 & 1 & 0 & 0 & \kappa\sqrt{\tau}\\
    -\mu\tau & 0 & 1 & 0 & 0\\
    0 & 0 & \kappa\sqrt{\tau} & 1 & 0\\
    0 & 0 & 0 & 0 & 1
  \end{pmatrix}.
\end{gather}
The initial covariance matrix is $\gamma_0 = \diag[ 2\var(B_0), 1, 1,
1, 1]$. After application of the matrix $\mathbf{S}$, and the
operations~\eqref{eq:19} and~\eqref{eq:21} representing the
polarization detection of the optical field, corresponding to a
homodyne detection of the variable $x_\text{ph}$, the atomic variables
and the B-field become correlated, a new beam segment enters in the
standard coherent state as described by~\eqref{eq:22}
and~\eqref{eq:13}, and the evolution proceeds. In the limit of short
beam segments, the evolution can be replaced by a Riccati differential
equation for the $3\times 3$ covariance matrix for the atoms and the
B-field, and this equation can be solved
analytically~\cite{molmer04:_estim_param_gauss_probes}. The variance
of the B-field is
\begin{gather}
  \label{eq:27}
  \begin{split}
    \var(B(t))
    &= \frac{\var(B_0)(\kappa^2t+1)}{
      \frac{1}{6}\kappa^4\mu^2\var(B_0)t^4
      + \frac{2}{3}\kappa^2\mu^2\var(B_0)t^3
      + \kappa^2t + 1}\\
    &\xrightarrow[t\to\infty]{}
    \frac{6}{\kappa^2\mu^2t^3}
    \propto \frac{1}{N_\text{at}^2\Phi t^3}
  \end{split}
\end{gather}
which yields precisely the error on the estimate of the B-field. We
note that the uncertainty of the field strength decreases as
$1/(N_\text{at}t^{3/2})$ and not as $1/\sqrt{N_\text{at}t}$ as one
might expect from standard counting statistics arguments. This
improved precision is due to the squeezing of the atomic spin during the
probing process.

\subsection{Squeezed light}
\label{sec:squeezed-light}

In Refs.~\cite{molmer04:_estim_param_gauss_probes} we modelled the use
of squeezed light by introducing the squeezing parameter $r$ such that
Eq.~\eqref{eq:22} is replaced by
\begin{gather}
  \label{eq:57}
  \mathbf{B}_\gamma \mapsto
  \begin{pmatrix}
    1/r & 0\\
    0 & r
  \end{pmatrix},
\end{gather}
i.e., every beam segment enters the interaction in a squeezed state.
Going though the calculations we find that $\kappa^2$ should be
replaced with $\kappa^2r$ in Eq.~\eqref{eq:27}: the B-field estimate
is improved.

As noted in Ref.~\cite{molmer04:_estim_param_gauss_probes}, this
treatment of a squeezed beam, in the limit of small $\tau$, is only
valid if the squeezing bandwidth is infinite. The squeezing properties
of the beam from an OPO, however, only reveal themselves if a narrow
frequency component is selected, or if the field is integrated over
times longer than the inverse bandwidth of squeezing, which are
certainly longer than the infinitesimal $\tau$ employed in the
continuous limit, where the Riccati equation is solved.

The full probing may well take longer than the inverse bandwidth, and
one would hence expect that one still benefits from the squeezing in
this longer time limit. We shall verify this assumption by a
calculation in which we treat the probing with the field coming out of
our OPO cavity in the full Gaussian formalism.

\begin{figure}[htbp]
  \centering
  \includegraphics[width=7cm]{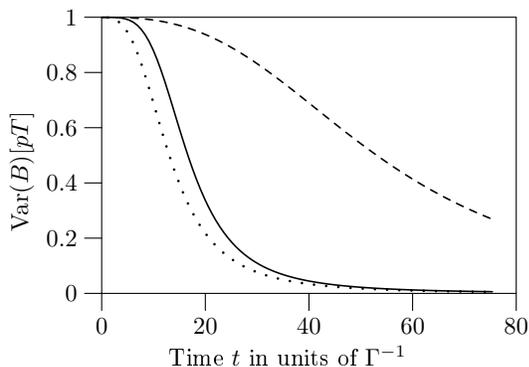}
  \caption{Variance of the $B$ field as a function of time. We use the
    same value of $g$ and $\Gamma$ as in Fig.~\ref{fig:4}, and
    $\kappa^2 = \SI{1.83e6}{s^{-1}}$ and $\mu =
    \SI{8.79e4}{(s~pT)^{-1}}$. The dashed line is without squeezing,
    the full line is with squeezed light generated in a cavity, and
    the dotted line is with the squeezing parameter $r$.}
  \label{fig:5}
\end{figure}

The example serves as a model for how to consider other atomic probing
schemes with realistic squeezed light sources. We treat as Gaussian
variables the B-field, the atomic variables, the intra-cavity field,
and a single segment of light $\mathbf{y} = (B_z, x_\text{at},
p_\text{at}, x_c, p_c, x_\text{ph}, p_\text{ph})^T$. The beam segment
enters on the cavity mirror in the vacuum state, it is reflected off
the mirror with some squeezing and some entanglement with the partly
transmitted intra-cavity field, it interacts with the atoms, and
finally it is detected by homodyne detection, causing a moderate
change of the joint covariance matrix for the B-field, the atoms, and
the intra-cavity field. The transformation to lowest order in $\tau$
of the variables is now given by
\begin{gather}
  \label{eq:28}
  \mathbf{S} =
  \begin{pmatrix}
    1 & 0 & 0 & 0 & 0 & 0 & 0\\
    \mu\tau & 1 & 0 & 0 & 0 & 0 & 0\\
    0 & 0 & 1 & \kappa\sqrt{\Gamma}\tau & 0 & -\kappa\sqrt{\tau} & 0\\
    0 & 0 & 0 & \xi+2g\tau & 0 & \sqrt{\Gamma\tau} & 0\\
    0 & 0 & 0 & 0 & \xi-2g\tau & 0 & \sqrt{\Gamma\tau}\\
    0 & 0 & 0 & -\sqrt{\Gamma\tau} & 0 & \xi & 0\\
    0 & -\kappa\sqrt{\tau} & 0 & 0 & -\sqrt{\Gamma\tau} & 0 & \xi
  \end{pmatrix},
\end{gather}
with $\xi = 1-\Gamma\tau/2$ as introduced in Eq.~\eqref{eq:43}. Note
that this matrix combines the elements present in the transformation
of the field components alone~\eqref{eq:43} and the B-field-atom and
light-atom interaction~\eqref{eq:25}. Again the beam segment is
inserted in its vacuum state~\eqref{eq:22}, and it is probed by
homodyne detection leading to the update formula~\eqref{eq:21}. The
bandwidth is taken care of by the intra-cavity field which establishes
the necessary correlation between beam segments detected at different
times. In the continuous limit we find the corresponding Riccati
equation, and its solution provides the variance of the B-field as a
function of time as shown in Fig.~\ref{fig:5}. The figure shows both
the results without squeezing, with finite bandwidth squeezing, and
the simple infinite bandwidth result~\eqref{eq:27} with a simple
squeezing parameter $r$ applied to each segment. We take the value $r
= \frac{(\Gamma+4g)^2}{(\Gamma-4g)^2}$, corresponding to the long-time
limit of Eq.~\eqref{eq:8}, and we see a good agreement for long times
between the two curves for squeezed states. We also see, that the
finite band-width curve is an improvement with respect to the case of
non-squeezed light, but that we have to probe for a certain time on
the order of the squeezing bandwidth before we see the effect of
squeezing. Indeed, the finite bandwidth curve is to a good
approximation simply delayed by $16g
\frac{3\Gamma+4g}{(\Gamma-4g)(\Gamma+4g)^2}$ compared with the
infinite broad-band squeezed light curve.

The analytical result for $\var(B)$ is very lengthy. For small times
$t$ we get the result without squeezing as can be seen in
Fig.~\ref{fig:5}, and for large $t$ the result is exactly the same as
in the infinite bandwidth case $\var(B(t)) = \frac{6}{\mu^2 r
  \kappa^2t^3}$ if we identify the squeezing parameter by $r =
\frac{(\Gamma+4g)^2}{(\Gamma-4g)^2}$.

\section{Conclusion and outlook}
\label{sec:outlook}

In summary, we have presented a Gaussian state description of the
light from an optical parametric oscillator and its interaction with
large atomic samples. The treatment is very effective, because the
state of the parts of the beam that have just left the OPO cavity can
be treated as a single mode, corresponding to a short beam segment,
and after the interaction, the segment can be eliminated from the
formalism. Here, we presented the dynamics when the field is probed by
homodyne detection, and it is turned into classical information; if
the beam propagates away without detection, it may be traced out of
the formalism, which is an even simpler operation in the Gaussian
formalism, since the corresponding rows and columns in the covariance
matrix should just be removed. Finite bandwidth effects are included
in the treatment by retaining the quantum state of the intra-cavity
field, which is also a single field mode, i.e., at the price of adding
a single pair of canonically conjugate variables ($x_c$, $p_c$), which
in the Gaussian formalism is done by adding two extra rows and columns
to the covariance matrix.

The use of squeezed light holds the potential to improve spin
squeezing, entanglement, and precision probing, and we demonstrated
such an improvement in the case of magnetometry compared with the
infinite bandwidth case, we also showed how
the finite bandwidth of squeezing manifests itself as a time lag
before the improvement is obtained in agreement with the observation
that squeezing is only present in a light beam, if one integrates a
sufficiently long part of the beam.

The method described is fully general, and further studies can be
carried out along the same lines on other proposals involving squeezed
light. It is readily generalized to incorporate more atomic systems,
more field modes, non-degenerate OPOs, and as we showed also finite
detection bandwidth can be modelled by the addition of auxiliary
modes. Finite bandwidth of the light sources and of the detection
system may also play non-trivial roles in conjunction with decay and
decoherence which set an upper limit to the degree of entanglement
obtained in gasses~\cite{sherson05:_entan_large_atomic_sampl}.

Finally we note that squeezed light has been proposed as an ingredient
in various quantum information protocols, such as
teleportation~\cite{furusawa98:_uncon_quant_telep}, as a source of
heralded single
photons~\cite{hong86:_exper_realiz_local_one_photon_state,myers04:_singl_photon_sourc_with_linear},
as a resource in continuous variable quantum computing and error
correction~\cite{braunstein98:_error_correc_contin_quant_variab}. In
many of these protocols, an elementary analysis is given in terms of
single mode fields, where indeed, a full time and frequency dependent
analysis would be more appropriate.

\begin{acknowledgments}
  LBM is supported by the Danish Natural Science Research Council (Grant
No.~21-03-0163).
\end{acknowledgments}

\appendix

\section{Squeezing of Output Field}
\label{sec:sque-outp-field}

To calculate the variance of $x_T$ and $p_T$ defined in
Eq.~\eqref{eq:32} we use the Gaussian description.
\begin{figure}[htbp]
  \centering
  \includegraphics{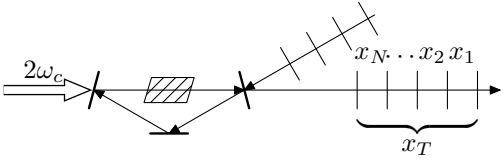}
  \caption{The figure shows how we label the beam segments which are
    accumulated in $x_T$. We only obtain squeezing if we observe many
    beam segments, not if we only observe one segment.}
  \label{fig:7}
\end{figure}

The initial variables are $\mathbf{y}_1 = (x_\text{c}, p_\text{c},
x_{\text{ph}_1}, p_{\text{ph}_1})$ and the initial intra-cavity field
and incident vacuum segment covariance matrix is
\begin{gather}
  \label{eq:2}
  \boldsymbol{\gamma}_1 =
  \begin{pmatrix}
    a_{11} & a_{12} & 0 & 0\\
    a_{12} & a_{22} & 0 & 0\\
    0 & 0 & 1 & 0\\
    0 & 0 & 0 & 1
  \end{pmatrix}.
\end{gather}
The associated transformation matrix is given by Eq.~\eqref{eq:43}
\begin{gather}
  \label{eq:3}
  \mathbf{S}_1 =
  \begin{pmatrix}
    \xi+2g\tau & 0 & \sqrt{\Gamma\tau} & 0\\
    0 & \xi-2g\tau & 0 & \sqrt{\Gamma\tau}\\
    -\sqrt{\Gamma\tau} & 0 & \xi & 0\\
    0 & -\sqrt{\Gamma\tau} & 0 & \xi
  \end{pmatrix}.
\end{gather}
where $\xi = 1-\Gamma\tau/2$ as introduced in Eq.~\eqref{eq:43}.
After the interaction $\tilde{\boldsymbol{\gamma}}_1 =
\textbf{S}_1\boldsymbol{\gamma}_1\mathbf{S}_1^T$. We now build the
dynamics recursively by inserting two rows and columns between the
second and third row and column in $\tilde{\boldsymbol{\gamma}}$. In
this way we represent the subsequent incident vacuum segments by
\begin{gather}
  \label{eq:4}
  \boldsymbol{\gamma}_{k+1} =
  \begin{pmatrix}
    \{\tilde{\boldsymbol{\gamma}}_k\}_{(1:2,1:2)}
    & 0 &
    \{\tilde{\boldsymbol{\gamma}}_k\}_{(1:2,3:2k+2)}\\
    0 & \mathbbm{1}_{2\times2} & 0\\
    \{\tilde{\boldsymbol{\gamma}}_k\}_{(3:2k+2,1:2)}
    & 0 &
    \{\tilde{\boldsymbol{\gamma}}_k\}_{(3:2k+2,3:2k+2)}
  \end{pmatrix}.
\end{gather}
The transformation matrix is
\begin{gather}
  \label{eq:5}
  \mathbf{S}_k =
  \begin{pmatrix}
    \mathbf{S}_1 & 0\\
    0 & \mathbbm{1}_{(2k-2)\times(2k-2)}
  \end{pmatrix},
\end{gather}
and $\tilde{\boldsymbol{\gamma}}_k = \mathbf{S}_k
\boldsymbol{\gamma}_k \mathbf{S}_k^T$. Eqs.~(\ref{eq:5},\ref{eq:4})
are now inserted, and the number of variables grows with time as we
get more and more light segments, $\mathbf{y}_N = (x_c, p_c,
x_{\text{ph}_N}, p_{\text{ph}_N}, \dots, x_{\text{ph}_1},
p_{\text{ph}_1})$

If $a_{12}=0$ then every second element in $\boldsymbol{\gamma}$ is
zero and $\boldsymbol{\gamma}$ can be rewritten on block diagonal form
with similar $x_\text{ph}$ and $p_\text{ph}$ blocks. The system of
equations for the $x_\text{ph}$, $\mathbf{y}_N = (x_c,
x_{\text{ph}_1}, \dots, x_{\text{ph}_N})$, variables can be written as
\begin{gather}
  \label{eq:63}
  \mathbf{S}_k =
  \begin{pmatrix}
    1-\Gamma\tau/2+2g\tau & \sqrt{\Gamma\tau} & 0\\
    -\sqrt{\Gamma\tau} & 1-\Gamma\tau/2 & 0\\
    0 & 0 & \mathbbm{1}_{(k-1)\times(k-1)}
  \end{pmatrix}
\end{gather}
\begin{gather}
  \label{eq:61}
  \boldsymbol{\gamma}_1 =
  \begin{pmatrix}
    a_{11} & 0\\
    0 & 1
  \end{pmatrix}
\end{gather}
\begin{gather}
  \label{eq:62}
  \boldsymbol{\gamma}_k =
  \begin{pmatrix}
    A_k & 0 & \mathbf{C}_k\\
    0 & 1 & 0\\
    \mathbf{C}_k^T & 0 & \mathbf{B}_k
  \end{pmatrix}
\end{gather}
where $A_k$ is a real number, $\mathbf{C}_k$ is a $1\times(k-1)$ row
vector, and $\mathbf{B}_k$ is a $(k-1)\times(k-1)$ matrix.

From this we find the recurrence equations
\begin{align}
  \label{eq:64}
  A_{k+1} &= (1-\Gamma\tau/2+2g\tau)^2 A_k + \Gamma\tau\\
  \mathbf{C}_{k+1}^T &=
  \begin{pmatrix}
    -\sqrt{\Gamma\tau}(1-\Gamma\tau/2+2g\tau)A_k + \sqrt{\Gamma\tau}(1-\Gamma\tau/2)\\
    (1-\Gamma\tau/2+2g\tau)\mathbf{C}_k^T
  \end{pmatrix}\\
  \mathbf{B}_{k+1} &=
  \begin{pmatrix}
    \Gamma\tau A_k + (1-\Gamma\tau/2)^2 & -\sqrt{\Gamma\tau}\mathbf{C}_k\\
    -\sqrt{\Gamma\tau}\mathbf{C}_k^T & \mathbf{B}_k
  \end{pmatrix}
\end{align}
which can be solved, and the variance of $x_T$ is found to be
\begin{gather}
  \label{eq:6}
  \begin{split}
    \var(x_T)
    &= \frac{1}{N} \sum_{i=1}^N \sum_{j=1}^N \cov(x_i,x_j)\\
    &=a_{11} \left[ \frac{\Gamma\tau}{2N}
      \frac{1-\alpha^{2N}}{1-\alpha^2}\right.\\
    &\hphantom{= a_{11} \Big[} \left.
      + \frac{\Gamma\tau\alpha^4}{N(1-\alpha)}
      \left( \frac{1-\alpha^N}{1-\alpha}
        - \frac{1-\alpha^{2N}}{1-\alpha^2} \right) \right]\\
    &\relphantom{=} +\frac{\Gamma\tau^2}{2(1-\alpha^N)}
    - \frac{\Gamma\tau^2}{2N} \frac{1-\alpha^{2N}}{(1-\alpha^2)^2}\\
    &\relphantom{=} + \tfrac{1}{2}(1-\Gamma\tau/2)^2
    - \Gamma\tau(1-\Gamma\tau/2)\frac{1}{1-\alpha}\\
    &\relphantom{=} + \frac{\Gamma\tau(1-\Gamma\tau/2)}{N}
    \frac{1-\alpha^N}{(1-\alpha)^2}\\
    &\relphantom{=} + \frac{\Gamma\tau^2\alpha}{N(1-\alpha^2)(1-\alpha)}
    \left( N - \frac{1-\alpha^N}{1-\alpha}\right.\\
    &\hphantom{= a_{11} \Big[} \left.
      + \alpha^3\frac{1-\alpha^{2N}}{1-\alpha^2} -
      \alpha^3\frac{1-\alpha^N}{1-\alpha} \right)
  \end{split}
\end{gather}
where $\alpha = 1-\Gamma\tau/2+2g\tau$. If we let $T=N\tau$ and then
let $\tau\to 0$ then
\begin{gather}
  \label{eq:7}
  \begin{split}
    \var(x_T)
    &\to \frac{1}{2T(\Gamma-4g)^3}
    \Bigl\{ (\Gamma-4g)(\Gamma+4g)^2T\\
    &\relphantom{\to} -4\Gamma(\Gamma+8g)
    + 4a_{11}\Gamma(\Gamma-4g)\\
    &\relphantom{\to} - 8\Gamma
    \bigl[ a_{11}(\Gamma-4g) - (\Gamma+4g)\bigr] e^{(-\Gamma/2+2g)T}\\
    &\relphantom{\to} - 4\Gamma \bigl[ \Gamma -
        a_{11}(\Gamma-4g)\bigr] e^{(-\Gamma+4g)T}\Bigr\}.
  \end{split}
\end{gather}
In the $T\to\infty$ limit, the first term dominates, and the
expression for $\var(x_T)$ does not depend upon $a_{11}$.

In steady state we may insert $a_{11} = \frac{\Gamma}{\Gamma-4g}$ in
Eq.~\eqref{eq:7}, and we obtain
\begin{gather}
  \label{eq:10}
  \begin{split}
    \var(x_T)
    &= \frac{1}{2T(\Gamma-4g)^3}
    \Bigl[ (\Gamma-4g)(\Gamma+4g)^2T\\
    &\relphantom{=} -32\Gamma g + 32\Gamma g e^{(-\Gamma/2+2g)T}\Bigr],
  \end{split}
\end{gather}
and
\begin{gather}
  \label{eq:58}
  \begin{split}
    \var(p_T)
    &= \frac{1}{2T(\Gamma+4g)^3}
    \Bigl[ (\Gamma+4g)(\Gamma-4g)^2T\\
    &\relphantom{=} - 32\Gamma g
    - 32g(\Gamma-4g)e^{-(\Gamma/2+2g)T}\\
    &\relphantom{=} - 64g^2e^{-(\Gamma+4g)T} \Bigr].
  \end{split}
\end{gather}
for the variances of the quadrature components.


\begin{thebibliography}{21}
\expandafter\ifx\csname natexlab\endcsname\relax\def\natexlab#1{#1}\fi
\expandafter\ifx\csname bibnamefont\endcsname\relax
  \def\bibnamefont#1{#1}\fi
\expandafter\ifx\csname bibfnamefont\endcsname\relax
  \def\bibfnamefont#1{#1}\fi
\expandafter\ifx\csname citenamefont\endcsname\relax
  \def\citenamefont#1{#1}\fi
\expandafter\ifx\csname url\endcsname\relax
  \def\url#1{\texttt{#1}}\fi
\expandafter\ifx\csname urlprefix\endcsname\relax\def\urlprefix{URL }\fi
\providecommand{\bibinfo}[2]{#2}
\providecommand{\eprint}[2][]{\url{#2}}

\bibitem[{\citenamefont{Fiur\'a\v{s}ek}(2002)}]{fiurasek02:_gauss_trans_distil%
_entan_gauss_states}
\bibinfo{author}{\bibfnamefont{J.}~\bibnamefont{Fiur\'a\v{s}ek}},
  \bibinfo{journal}{Phys. Rev. Lett.} \textbf{\bibinfo{volume}{89}},
  \bibinfo{pages}{137904} (\bibinfo{year}{2002}).

\bibitem[{\citenamefont{Giedke and
  Cirac}(2002)}]{giedke02:_charac_gauss_operat_distil_gauss_states}
\bibinfo{author}{\bibfnamefont{G.}~\bibnamefont{Giedke}} \bibnamefont{and}
  \bibinfo{author}{\bibfnamefont{J.~I.} \bibnamefont{Cirac}},
  \bibinfo{journal}{Phys. Rev. A} \textbf{\bibinfo{volume}{66}},
  \bibinfo{pages}{032316} (\bibinfo{year}{2002}).

\bibitem[{\citenamefont{Eisert and
  Plenio}(2003)}]{eisert03:_introd_basic_entan_theor_contin_variab_system2}
\bibinfo{author}{\bibfnamefont{J.}~\bibnamefont{Eisert}} \bibnamefont{and}
  \bibinfo{author}{\bibfnamefont{M.~B.} \bibnamefont{Plenio}},
  \bibinfo{journal}{Int.\ J.\ Quant.\ Inf.} \textbf{\bibinfo{volume}{1}},
  \bibinfo{pages}{479} (\bibinfo{year}{2003}).

\bibitem[{\citenamefont{Mølmer and
  Madsen}(2004)}]{molmer04:_estim_param_gauss_probes}
\bibinfo{author}{\bibfnamefont{K.}~\bibnamefont{Mølmer}} \bibnamefont{and}
  \bibinfo{author}{\bibfnamefont{L.~B.} \bibnamefont{Madsen}},
  \bibinfo{journal}{Phys. Rev. A} \textbf{\bibinfo{volume}{70}},
  \bibinfo{pages}{052102} (\bibinfo{year}{2004}).

\bibitem[{\citenamefont{Madsen and
  Mølmer}(2004)}]{madsen04:_spin_squeez_precis_probin_light2}
\bibinfo{author}{\bibfnamefont{L.~B.} \bibnamefont{Madsen}} \bibnamefont{and}
  \bibinfo{author}{\bibfnamefont{K.}~\bibnamefont{Mølmer}},
  \bibinfo{journal}{Phys. Rev. A} \textbf{\bibinfo{volume}{70}},
  \bibinfo{pages}{052324} (\bibinfo{year}{2004}).

\bibitem[{\citenamefont{Petersen et~al.}(2005)\citenamefont{Petersen, Madsen,
  and Mølmer}}]{petersen05:_magnet_entan_atomic_sampl}
\bibinfo{author}{\bibfnamefont{V.}~\bibnamefont{Petersen}},
  \bibinfo{author}{\bibfnamefont{L.~B.} \bibnamefont{Madsen}},
  \bibnamefont{and} \bibinfo{author}{\bibfnamefont{K.}~\bibnamefont{Mølmer}},
  \bibinfo{journal}{Phys. Rev. A} \textbf{\bibinfo{volume}{71}},
  \bibinfo{pages}{012312} (\bibinfo{year}{2005}).

\bibitem[{\citenamefont{Scully and Zubairy}(1997)}]{scully97:_quant_optic}
\bibinfo{author}{\bibfnamefont{M.~O.} \bibnamefont{Scully}} \bibnamefont{and}
  \bibinfo{author}{\bibfnamefont{M.~S.} \bibnamefont{Zubairy}},
  \emph{\bibinfo{title}{Quantum Optics}} (\bibinfo{publisher}{Cambridge
  University Press}, \bibinfo{year}{1997}).

\bibitem[{\citenamefont{Collett and
  Gardiner}(1984)}]{collett84:_squeez_intrac_travel_wave_light}
\bibinfo{author}{\bibfnamefont{M.~J.} \bibnamefont{Collett}} \bibnamefont{and}
  \bibinfo{author}{\bibfnamefont{C.~W.} \bibnamefont{Gardiner}},
  \bibinfo{journal}{Phys. Rev. A} \textbf{\bibinfo{volume}{30}},
  \bibinfo{pages}{1386} (\bibinfo{year}{1984}).

\bibitem[{\citenamefont{Gardiner}(1991)}]{gardiner91:_quant_noise}
\bibinfo{author}{\bibfnamefont{C.~W.} \bibnamefont{Gardiner}},
  \emph{\bibinfo{title}{Quantum Noise}} (\bibinfo{publisher}{Springer},
  \bibinfo{address}{Berlin, Heidelberg}, \bibinfo{year}{1991}).

\bibitem[{\citenamefont{Walls and Milburn}(1994)}]{walls94:_quant_optic}
\bibinfo{author}{\bibfnamefont{D.~F.} \bibnamefont{Walls}} \bibnamefont{and}
  \bibinfo{author}{\bibfnamefont{G.~J.} \bibnamefont{Milburn}},
  \emph{\bibinfo{title}{Quantum Optics}} (\bibinfo{publisher}{Springer-Verlag},
  \bibinfo{year}{1994}).

\bibitem[{\citenamefont{Stockton et~al.}(2004)\citenamefont{Stockton, Geremia,
  Doherty, and Mabuchi}}]{stockton04:_robus_quant_param_estim}
\bibinfo{author}{\bibfnamefont{J.~K.} \bibnamefont{Stockton}},
  \bibinfo{author}{\bibfnamefont{J.~M.} \bibnamefont{Geremia}},
  \bibinfo{author}{\bibfnamefont{A.~C.} \bibnamefont{Doherty}},
  \bibnamefont{and} \bibinfo{author}{\bibfnamefont{H.}~\bibnamefont{Mabuchi}},
  \bibinfo{journal}{Phys. Rev. A} \textbf{\bibinfo{volume}{69}},
  \bibinfo{pages}{032109} (\bibinfo{year}{2004}).

\bibitem[{\citenamefont{Sørensen}()}]{soerensen:_privat}
\bibinfo{author}{\bibfnamefont{J.~L.} \bibnamefont{Sørensen}},
  \emph{\bibinfo{title}{Private communication}}.

\bibitem[{\citenamefont{Geremia et~al.}(2003)\citenamefont{Geremia, Stockton,
  Doherty, and
  Mabuchi}}]{geramia03:_quant_kalman_filter_heisen_limit_atomic_magnet}
\bibinfo{author}{\bibfnamefont{J.~M.} \bibnamefont{Geremia}},
  \bibinfo{author}{\bibfnamefont{J.~K.} \bibnamefont{Stockton}},
  \bibinfo{author}{\bibfnamefont{A.~C.} \bibnamefont{Doherty}},
  \bibnamefont{and} \bibinfo{author}{\bibfnamefont{H.}~\bibnamefont{Mabuchi}},
  \bibinfo{journal}{Phys. Rev. Lett.} \textbf{\bibinfo{volume}{91}},
  \bibinfo{pages}{250801} (\bibinfo{year}{2003}).

\bibitem[{\citenamefont{Kuzmich, Mandel, Janis, Young, Ejnisman, and
  Bigelow}(1999)}]{kuzmich99:_quant_nondem_measur_collec_atomic_spin}
\bibinfo{author}{\bibfnamefont{A.}~\bibnamefont{Kuzmich}},
  \bibinfo{author}{\bibfnamefont{L.}~\bibnamefont{Mandel}},
  \bibinfo{author}{\bibfnamefont{J.}~\bibnamefont{Janis}},
  \bibinfo{author}{\bibfnamefont{Y.~E.} \bibnamefont{Young}},
  \bibinfo{author}{\bibfnamefont{R.}~\bibnamefont{Ejnisman}}, \bibnamefont{and}
  \bibinfo{author}{\bibfnamefont{N.~P.} \bibnamefont{Bigelow}},
  \bibinfo{journal}{Phys. Rev. A} \textbf{\bibinfo{volume}{60}},
  \bibinfo{pages}{2346} (\bibinfo{year}{1999}).

\bibitem[{\citenamefont{Julsgaard et~al.}(2001)\citenamefont{Julsgaard,
  Kozhekin, and Polzik}}]{julsgaard01:_exper_long_entan_two_macros_objec}
\bibinfo{author}{\bibfnamefont{B.}~\bibnamefont{Julsgaard}},
  \bibinfo{author}{\bibfnamefont{A.}~\bibnamefont{Kozhekin}}, \bibnamefont{and}
  \bibinfo{author}{\bibfnamefont{E.~S.} \bibnamefont{Polzik}},
  \bibinfo{journal}{Nature} \textbf{\bibinfo{volume}{413}},
  \bibinfo{pages}{400} (\bibinfo{year}{2001}).

\bibitem[{\citenamefont{Julsgaard et~al.}(2004)\citenamefont{Julsgaard,
  Sherson, Cirac, Fiur\'a\v{s}ek, and
  Polzik}}]{julsgaard04:_exper_demon_quant_memor_light}
\bibinfo{author}{\bibfnamefont{B.}~\bibnamefont{Julsgaard}},
  \bibinfo{author}{\bibfnamefont{J.}~\bibnamefont{Sherson}},
  \bibinfo{author}{\bibfnamefont{J.~I.} \bibnamefont{Cirac}},
  \bibinfo{author}{\bibfnamefont{J.}~\bibnamefont{Fiur\'a\v{s}ek}},
  \bibnamefont{and} \bibinfo{author}{\bibfnamefont{E.}~\bibnamefont{Polzik}},
  \bibinfo{journal}{Nature} \textbf{\bibinfo{volume}{432}},
  \bibinfo{pages}{482} (\bibinfo{year}{2004}).

\bibitem[{\citenamefont{Sherson and
  Mølmer}(2005)}]{sherson05:_entan_large_atomic_sampl}
\bibinfo{author}{\bibfnamefont{J.}~\bibnamefont{Sherson}} \bibnamefont{and}
  \bibinfo{author}{\bibfnamefont{K.}~\bibnamefont{Mølmer}},
  \bibinfo{journal}{Phys. Rev. A} \textbf{\bibinfo{volume}{71}},
  \bibinfo{pages}{033813} (\bibinfo{year}{2005}).

\bibitem[{\citenamefont{Furusawa et~al.}(1998)\citenamefont{Furusawa, Sørensen,
  Braunstein, Fuchs, Kimble, and Polzik}}]{furusawa98:_uncon_quant_telep}
\bibinfo{author}{\bibfnamefont{A.}~\bibnamefont{Furusawa}},
  \bibinfo{author}{\bibfnamefont{J.~L.} \bibnamefont{Sørensen}},
  \bibinfo{author}{\bibfnamefont{S.~L.} \bibnamefont{Braunstein}},
  \bibinfo{author}{\bibfnamefont{C.~A.} \bibnamefont{Fuchs}},
  \bibinfo{author}{\bibfnamefont{H.~J.} \bibnamefont{Kimble}},
  \bibnamefont{and} \bibinfo{author}{\bibfnamefont{E.~S.}
  \bibnamefont{Polzik}}, \bibinfo{journal}{Science}
  \textbf{\bibinfo{volume}{282}}, \bibinfo{pages}{706} (\bibinfo{year}{1998}).

\bibitem[{\citenamefont{Hong and
  Mandel}(1986)}]{hong86:_exper_realiz_local_one_photon_state}
\bibinfo{author}{\bibfnamefont{C.~K.} \bibnamefont{Hong}} \bibnamefont{and}
  \bibinfo{author}{\bibfnamefont{L.}~\bibnamefont{Mandel}},
  \bibinfo{journal}{Phys. Rev. Lett.} \textbf{\bibinfo{volume}{56}},
  \bibinfo{pages}{58} (\bibinfo{year}{1986}).

\bibitem[{\citenamefont{Myers et~al.}(2004)\citenamefont{Myers, Ericsson, and
  Laflamme}}]{myers04:_singl_photon_sourc_with_linear}
\bibinfo{author}{\bibfnamefont{C.~R.} \bibnamefont{Myers}},
  \bibinfo{author}{\bibfnamefont{M.}~\bibnamefont{Ericsson}}, \bibnamefont{and}
  \bibinfo{author}{\bibfnamefont{R.}~\bibnamefont{Laflamme}},
  \emph{\bibinfo{title}{A single photon source with linear optics and squeezed
  states}}, \bibinfo{howpublished}{quant-ph/0408194} (\bibinfo{year}{2004}).

\bibitem[{\citenamefont{Braunstein}(1998)}]{braunstein98:_error_correc_contin_%
quant_variab}
\bibinfo{author}{\bibfnamefont{S.~L.} \bibnamefont{Braunstein}},
  \bibinfo{journal}{Phys. Rev. Lett.} \textbf{\bibinfo{volume}{80}},
  \bibinfo{pages}{4084} (\bibinfo{year}{1998}).

\end{thebibliography}

\end{document}